\documentclass[10pt, conference]{IEEEtran}
\IEEEoverridecommandlockouts
\usepackage{cite}
\usepackage{amsmath,amssymb,amsfonts}
\usepackage{algorithmic}
\usepackage{graphicx}
\usepackage{textcomp}
\usepackage{xcolor}
\usepackage[hidelinks,hypertexnames=false]{hyperref}
\usepackage{float}
\usepackage{booktabs}

\def\BibTeX{{\rm B\kern-.05em{\sc i\kern-.025em b}\kern-.08em
    T\kern-.1667em\lower.7ex\hbox{E}\kern-.125emX}}

\newcommand{\ZenodoDOI}{\href{https://doi.org/10.5281/zenodo.19778806}{doi:10.5281/zenodo.19778806}}
    
\begin{document}

\title{Topological State-Aware Simulation Framework for Inter-Satellite Twin-Field QKD Networks\\
% {\footnotesize \textsuperscript{*}Note: Sub-titles are not captured in Xplore and should not be used}
}

\author{\IEEEauthorblockN{1\textsuperscript{st} Sergio Vázquez-Pozo}
\IEEEauthorblockA{\textit{Software Engineering Undergraduate} \\
\textit{University of Extremadura}\\
Cáceres, Spain \\
svazquezzp@gmail.com \\
orcid.org/0009-0006-7500-5959}
\and
\IEEEauthorblockN{2\textsuperscript{nd} Juan Manuel Murillo}
\IEEEauthorblockA{\textit{Quercus Software Engineering Group} \\
\textit{University of Extremadura}\\
Cáceres, Spain \\
juanmamu@unex.es \\
orcid.org/0000-0003-4961-4030}
}

\maketitle

\begin{abstract}
Inter-satellite links (ISLs) are the mandatory backbone for global quantum networks. While Twin-Field Quantum Key Distribution (TF-QKD) successfully surpasses linear rate-loss bounds, its extreme phase sensitivity makes it highly vulnerable to dynamic, non-IID (Independent and Identically Distributed) orbital environments. In composable finite-key analyses governed by the Generalized Entropy Accumulation Theorem (GEAT), traditional adaptive post-selection heuristics either violate strict independence conditions or incur massive second-order penalties that collapse the secret key rate. To overcome this, we introduce a reference-only topological post-selection oracle. By modeling the constellation as a Cellular Sheaf and applying Topological Data Analysis (TDA), our protocol derives a public acceptance event ($\Omega$) exclusively from classical beacon telemetry. To rigorously validate this mechanism, we develop a modular simulation framework equipped with stochastic noise injection and an explicit GEAT security ledger. Simulations across 2,000--5,000 km ISL separations compare the same Hodge--Koopman gate with TDA disabled and enabled. At 2,000 km, the median conditional candidate rates are $2.14\times10^{-6}$ and $5.87\times10^{-7}$ bit per emitted pulse, respectively; both configurations return zero at 3,000--5,000 km. TDA is active in all 4,788 evaluated windows, but does not extend the positive-candidate range in this scenario. These exported rates are conditional numerical candidates: the full protocol-level composable-security proof remains incomplete and the certified composable rate is therefore zero throughout.\end{abstract}


\section{Introduction}
\label{sec:intro}

The realization of a global Quantum Internet relies critically on overcoming the exponential photon loss that fundamentally bounds terrestrial optical fiber links. To achieve global-scale entanglement distribution, Satellite Quantum Key Distribution (Sat-QKD) provides the mandatory space-based backbone. Within this architecture, Inter-Satellite Links (ISL) are paramount, acting as the dynamic bridges that bypass atmospheric turbulence to establish a continuous, fiber-less global quantum network \cite{Liao2017Satellite, Yin2020Satellite, Pirandola2021FreeSpaceLimits, Vergoossen2019SatConstellations}.

Point-to-point QKD protocols are strictly constrained by the Pirandola-Laurenza-Ottaviani-Banchi (PLOB) bound \cite{PirandolaPLOB17}, which dictates the maximal secure communication rate over a lossy channel. Twin-Field QKD (TF-QKD) successfully overcomes this fundamental limit by relying on single-photon interference at an intermediate, untrusted herald node, effectively doubling the secure transmission distance in optical fibers \cite{Lucamarini18, MaSNS18, Curty19, Liu2023TF1000km} and emerging free-space architectures \cite{Li2025FreeSpaceTFQKD}. However, deploying phase-sensitive protocols like TF-QKD over ISLs introduces formidable operational challenges \cite{Cao2020FreeSpaceMDIQKD}. Because single-photon interference requires sub-wavelength phase stabilization between distant moving platforms, the protocol is highly susceptible to rapid phase-noise variations driven by relative orbital velocities, satellite platform jitter, and continuously reconfiguring spatial topologies.

Given the prohibitive costs and logistical complexity of aerospace deployments, high-fidelity simulation frameworks are the only viable path to certify protocol robustness and predict operational viability prior to hardware construction. However, current quantum network simulators often lack the integrated architecture required to orchestrate these intersecting domains. Existing tools typically isolate the physical layer simulation from the rigorous, composable security modeling required in finite-size, non-IID (Independent and Identically Distributed) regimes. This decoupling leads to over-optimistic secret key rate estimations and potentially compromises the cryptographic integrity of the system by failing to account for real-world orbital correlations. For example, most prior art assumes static point-to-point links and relies on empirical, ad-hoc phase-variance cut-offs for data post-selection \cite{Shan2024SNSPhasePostselection}. This approach is highly problematic in dynamic ISL deployments: without tracking the global topological consistency of the constellation, post-selection heuristics introduce severe methodological side-channels. If the topological state dynamics are ignored, the Generalized Entropy Accumulation Theorem (GEAT) dictates massive second-order finite-size penalties that collapse the secret key rate to zero \cite{MetgerRenner2022QKDGEA, Metger2022GEA}.

To bridge this gap, this paper introduces two main contributions. The primary contribution is the introduction of a novel Topological Protocol based on Cellular Sheaves \cite{HansenGhrist2019CellularSheaves}, integrated explicitly with advanced Topological Data Analysis (TDA) \cite{Carlsson2009TDA}. We model the ISL constellation as a cellular complex and expose synchronization errors through Hodge Laplacians. This mechanism functions as a rigorous, decentralized state oracle, ensuring that data post-selection is strictly derived from public reference telemetry without relying on a single central node, effectively isolating quantum events from network topology decisions. Consequently, this topological protocol structurally prevents the security ledger from misinterpreting dynamic orbital noise as adversarial interference, preserving the secret key rate.

To empirically prove this protocol against the strict constraints of finite-size entropy penalizations \cite{MetgerRenner2022QKDGEA}, we introduce our secondary contribution: a standardized, Composable Space-Network Simulation Framework. Because existing tools could not jointly process fluctuating orbital dynamics and strict Generalized Entropy Accumulation Theorem (GEAT) ledgers, this custom framework was required to evaluate the protocol. In addition, the resulting open-source simulation architecture stands as a powerful tool for the community, allowing network architects to evaluate rigorous composable secret key rates under realistic orbital constraints prior to prohibitive hardware deployments.

In accordance with open-science principles, we release the complete simulation codebase, configuration parameters, and generated artifacts (plots, JSON exports, and run metadata) via Zenodo: \ZenodoDOI.

The remainder of this paper is organized as follows. Section \ref{sec:channel_model} defines the essential background context: the PLOB bounds, dynamic ISL channel models, and the established structural challenges of TF-QKD in space. Section \ref{sec:topological_protocol} details our primary contribution—the proposed topological protocol via Cellular Sheaves. Section \ref{sec:sim_architecture} explains the resulting modular architecture of the simulation framework. Section \ref{sec:evaluation} presents empirical interpretations and a discussion of the trade-offs. Finally, Section \ref{sec:conclusions} summarises our conclusions.

\section{Background: ISL Channel Modeling and Limitations}
\label{sec:channel_model}

In this section, we outline the fundamental physical constraints and stochastic channel models that govern Inter-Satellite Links (ISLs), establishing the necessity for our topological approach. First, we review the fundamental bounds of repeaterless quantum communication to justify the selection of the TF-QKD architecture (Subsec. \ref{subsec:plob}). Next, we detail the ISL link budget, diffraction-limited coupling, and stochastic pointing jitter that define the physical substrate (Subsecs. \ref{subsec:islbudget} to \ref{subsec:physical_imperfections}). Finally, we model the phase drift and Hidden Markov Model (HMM) block dynamics that characterize the non-IID nature of the channel, illustrating the structural flaws of traditional post-selection mechanisms and motivating our state-aware protocol (Subsecs. \ref{subsec:gaussian_drift} and \ref{subsec:hmm_postselect}).  

\subsection{The PLOB Bound and Baseline Attenuations}
\label{subsec:plob}
The establishment of long-distance quantum links must confront the inescapable reality of photon loss. In any point-to-point QKD protocol acting over a pure-loss channel, the maximum secret key rate $R$ is dictated by the Pirandola-Laurenza-Ottaviani-Banchi (PLOB) bound, remaining strictly constrained by $R \le -\log_2(1-\eta)$ where $\eta$ is the channel transmittance \cite{PirandolaPLOB17}. To overcome this severe linear attenuation scaling, Twin-Field (TF) architectures \cite{Lucamarini18} intercede a central untrusted herald station to achieve an $O(\sqrt{\eta})$ scaling, vastly extending the terrestrial reach \cite{Chen2021TF511km, Liu2023TF1000km}. Understanding this theoretical limit establishes the baseline for our framework: achieving viable key rates over intercontinental ISL distances dictates the use of TF-QKD. However, because this architectural choice relies on single-photon interference, it introduces phase-sensitivity and vulnerability to orbital dynamics that our proposed topological oracle is designed to mitigate.

\subsection{ISL Link Budget Decomposition}
\label{subsec:islbudget}
For a vacuum ISL, attenuation is dominated by diffraction and terminal coupling rather than atmospheric extinction. We decompose the instantaneous transmittance as
\begin{equation}
    \eta(t)=\eta_{\text{tx}}\,\eta_{\text{rx}}(L(t))\,\eta_{\text{sys}}\,\eta_{\text{ptr}}(t),
\end{equation}
where $\eta_{\text{tx}}$ models transmitter optical throughput, $\eta_{\text{rx}}(L)$ is the diffraction-limited geometric capture efficiency at range $L$, $\eta_{\text{sys}}$ aggregates internal receiver efficiency (detectors, coupling, filtering), and $\eta_{\text{ptr}}$ is a stochastic pointing-induced penalty. For the TF geometry with an intermediate interferometer, we distinguish the direct (one-shot) end-to-end transmittance $\eta_{\text{dir}}(L)$ from the two-arm nominal TF channel $\eta_{\text{TF}}(L)\approx \eta_{\text{arm}}(L/2)^2$. Explicitly isolating these variables is fundamental for our framework, as the simulation must generate accurate, multi-layered telemetry to feed the topological oracle without conflating deterministic distance losses with stochastic anomalies. 

\subsection{Diffraction-Limited Geometric Coupling}
\label{subsec:geometric_coupling}
We assume a Gaussian beam of waist $w_0$ at the transmitter. The beam radius at the receiver plane is
\begin{equation}
    w(L)=w_0\sqrt{1+\left(\frac{L}{z_R}\right)^2},\qquad z_R=\frac{\pi w_0^2}{\lambda},
\end{equation}
with wavelength $\lambda$. For a circular receiver aperture of radius $r_{\text{rx}}=D_{\text{rx}}/2$, the geometric capture efficiency is approximated by the encircled-energy fraction
\begin{equation}
    \eta_{\text{rx}}(L)=1-\exp\left(-2\,\frac{r_{\text{rx}}^2}{w(L)^2}\right),
\end{equation}
which is the dominant deterministic loss term in deep-space ISLs \cite{AndrewsPhillips2005}. Integrating this baseline into our orbital module ensures that the topological protocol evaluates true signal anomalies rather than predictable geometric expansions during orbital passes.

\subsection{Pointing Jitter, Beam Wander, and Block Fading}
\label{subsec:physical_imperfections}
In practice, micro-vibrations and imperfect fine-steering control induce angular jitter that displaces the beam spot relative to the receiver aperture. Rather than attempting pulse-level integration of the closed-loop attitude dynamics, we model the residual penalty $\eta_{\text{ptr}}(t)$ as a multiplicative random process that yields block-wise log-normal fades, consistent with standard free-space optical link modeling \cite{Yin2020Satellite, AndrewsPhillips2005, Pirandola2021FreeSpaceLimits}. This is the primary mechanism by which the ISL deviates from the IID idealization: the click probabilities and phase-error proxies become correlated in time. These temporal correlations represent the exact non-IID behavior that triggers GEAT penalizations in standard frameworks. Therefore, accurately modeling this fading mechanism is what allows us to properly demonstrate the mitigating capabilities of our topological oracle.

\subsection{Gauss--Markov Phase Drift and Control Time-Scales}
\label{subsec:gaussian_drift}
TF-QKD is phase sensitive. We model the public residual phase drift (observable from strong reference telemetry) as a Gauss--Markov (AR(1)) process at a control sampling cadence $\Delta t_{\text{ctrl}}$:
\begin{equation}
    \phi_{k+1}=\rho_{\phi}\,\phi_k+\sigma_{\text{step}}\,\xi_k,\qquad \xi_k\sim\mathcal{N}(0,1),
\end{equation}
where $\rho_{\phi}\in(0,1)$ captures the coherence time of the platform/optical path and $\sigma_{\text{step}}$ sets the stationary variance. This explicitly yields non-IID blocks when aggregated over block intervals $\Delta t_{\text{block}}\gg\Delta t_{\text{ctrl}}$. Because the proposed topological protocol identifies synchronization loss by tracking continuous drift across the constellation, generating a realistic, temporally correlated phase model is a prerequisite for the simulation framework.

\subsection{HMM Block Dynamics and the Post-Selection Problem}
\label{subsec:hmm_postselect}
To simulate orbital variability without prohibitive sub-pulse integration, literature standardizes a Hidden Markov Model (HMM) or block-state approach \cite{Rabiner1989HMM}. The orbit is discretized into temporal blocks, each characterized by latent states for fading and phase drift. In our simulator, both the (log-)fading and the phase latent states can be instantiated as correlated AR(1) processes with parameters $(\rho_{\text{fading}},\rho_{\phi})$ and then sampled into block-wise effective observables.

Standard TF-QKD post-selection heuristics employ moving average cutoffs to isolate "bad" blocks \cite{Shan2024SNSPhasePostselection, Zhou2023PhaseErrorPostselection}. However, in finite-key limits under severe LEO (Low-Earth Orbit) transitions and Doppler-driven phase slips, relying solely on empirical cutoffs is dangerous: adaptive acceptance can correlate with private key-generation variables, and the GEAT/EAT machinery \cite{ArnonFriedman18EAT} then assigns worst-case second-order corrections that collapse the composable key rate \cite{MetgerRenner2022QKDGEA, Kamin2024GEATDecoy}. This establishes the requirement for our protocol to observe \emph{global} structural consistency of the constellation using public telemetry only.

Unlike fixed terrestrial fiber links, the validity and quality of the active inter-satellite links, denoted as the edge set $E(t)$ in our dynamic network graph, depend strictly on the instantaneous line-of-sight and the underlying relative Doppler velocity. Traditional QKD protocols abort unconditionally when the QBER spikes. For an ISL network to survive, it must embrace state-aware post-selection: discarding severely misaligned temporal blocks while harvesting keys during periods of optimal geometrical stability, without inducing cryptographic leakages. This HMM-based discretization provides the formal temporal structure for our framework, allowing the topological oracle to evaluate the structural consistency of the constellation and resolve the post-selection problem through a rigorous, non-adaptive gating mechanism.

\section{Topological Protocol via Cellular Sheaves}
\label{sec:topological_protocol}

The core deficiency in existing multi-node TF-QKD implementations is the reliance on isolated, per-link empirical thresholds. When phase reference frames drift across multiple satellites seamlessly, evaluating a single $A \to B$ link is insufficient and fragile. To robustly monitor the ISL network without a single point of failure or centralized trust, we introduce our primary contribution: a continuous topological protocol that evaluates network-wide synchronization consistency via sheaves.

In this section, we define the mathematical and operational structure of this protocol. First, we formalize the dynamic satellite network as a Cellular Sheaf (Subsec. \ref{subsec:sheaf_construction}). Next, we design a Topological Data Analysis (TDA) pipeline to extract persistent homology metrics from public telemetry (Subsec. \ref{subsec:tda_pipeline}). We then introduce the discrete Hodge decomposition to isolate synchronization obstructions (Subsec. \ref{subsec:hodge_oracle}), and finally, we formalize the public acceptance event that securely interfaces with the finite-key ledger (Subsec. \ref{subsec:public_event}).

\subsection{Cellular Sheaf Construction ($G(V, E, t)$)}
\label{subsec:sheaf_construction}
Instead of treating the ISL constellation as a standard moving attenuation segment, we promote the dynamic geometric constellation network $G = (V, E, t)$ to a Topological Cellular Complex $X$. Nodes $v \in V$ correspond to 0-cells and active spatial optical links $e \in E$ to 1-cells. To capture the physical state dynamics, we equip this space with a Cellular Sheaf $\mathcal{F}$ \cite{Curry2014Sheaves, HansenGhrist2019CellularSheaves}.

A Sheaf $\mathcal{F}$ assigns a vector space $F(\sigma)$ (the \textit{stalk}) to every cell $\sigma \in X$, mapping the local physical telemetry observed by the satellite. Let $x_v(t) \in \mathbb{R}^d$ be the continuous public reference signals generated at node $v$ (e.g., local oscillator phase drift estimators, pointing tracking angles). For an incident edge $v \triangleleft e$ (i.e., node $v$ connects via link $e$), the sheaf dictates a linear restriction map $\mathcal{F}_{v \triangleleft e}: F(v) \to F(e)$.

In an ideal, noiseless ISL topology, the restricted signals from two connected nodes must align within the edge's reference frame $F(e)$:
\begin{equation}
    \mathcal{F}_{v \triangleleft e} x_v = \mathcal{F}_{u \triangleleft e} x_u \quad \forall e=(u,v) \in E
\end{equation}
Any failure in this equation physically translates into broken coherence at the intermediate TF-QKD beam splitter, predicting an imminent collapse of the quantum bit error rate.

\paragraph{Optical Bundle Sub-Complex}
To make this physical, consider the specific optical bundle sub-complex at the TF-QKD herald node (usually the central satellite $C$). The stalk $F(C)$ represents the interference reference frame. The restriction maps $\mathcal{F}_{A \triangleleft e}$ and $\mathcal{F}_{B \triangleleft e}$ from Alice and Bob mathematically encode the path-length deformations and Doppler-induced phase shifts accumulated along the channels. If the combined telemetry restrictions into $F(C)$ do not commute perfectly, the sheaf mathematically exposes a non-zero homology, signaling that the optical bundle has lost its phase coherence. Formulating the protocol through this sub-complex allows the system to identify coherence losses across the intermediate node, providing a metric that isolated link evaluations cannot capture.

\begin{figure}[htbp]
    \centering
      \includegraphics[width=0.82\linewidth]{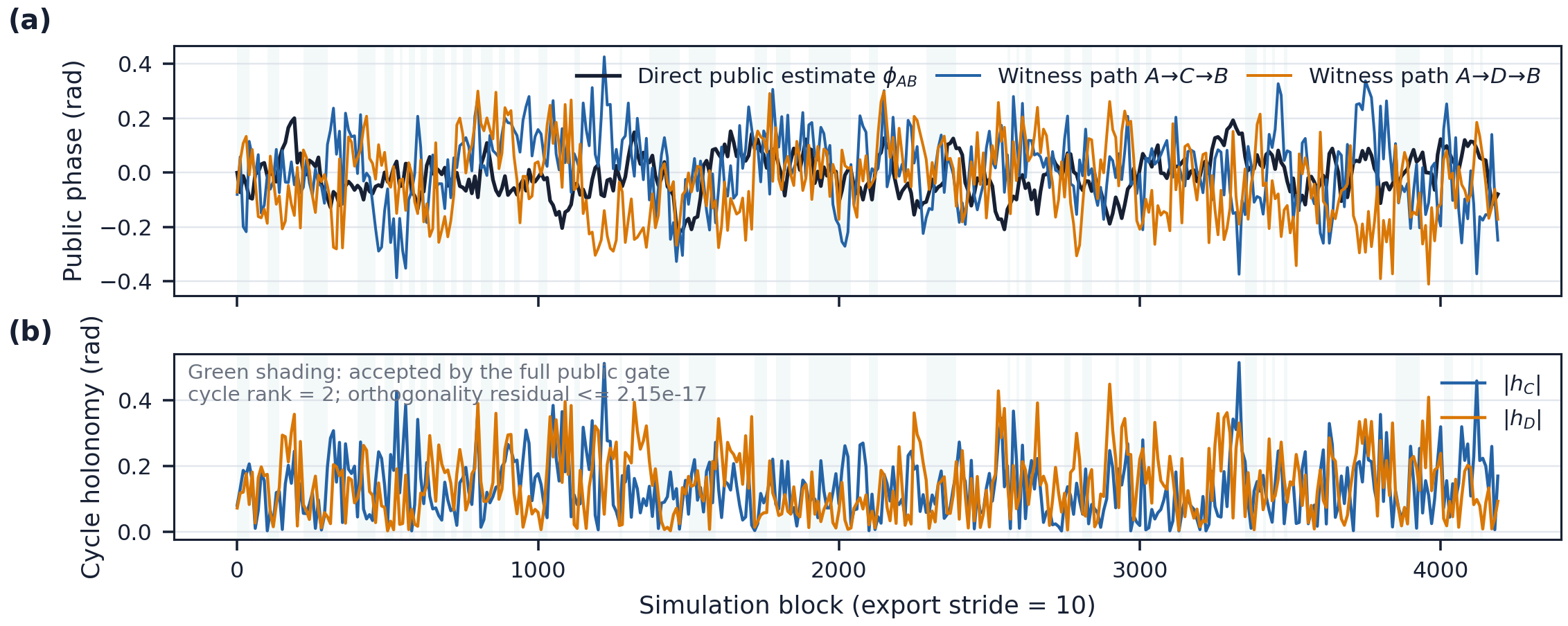}%    
    \caption{Exported public telemetry for the direct link and the two witness paths, together with their cycle holonomies. Green shading marks blocks accepted by the full public gate.}
    \label{fig:sheaf_bundle}
\end{figure}

\subsection{Topological Data Analysis (TDA) Pipeline}
\label{subsec:tda_pipeline}
To translate the abstract sheaf into a computable metric, we formalize a Topological Data Analysis (TDA) pipeline on the telemetry streams. The sequence is defined as $S \to K \to \mathcal{X} \to D_{k} \to q \to T$, bridging raw public signals to an auditable acceptance event via topological summaries. Here, $S$ denotes beacon/reference observables, $K$ is a public-only phase estimator used to isolate residual phase, $\mathcal{X}$ is the induced telemetry point cloud, $D_k$ are persistence descriptors, and $q$ is a scalar TDA consistency score \cite{Zomorodian2005, Carlsson2009TDA, CohenSteiner2007}.

First, standard linear estimators fail against the compound chaotic oscillation of LEO platform jitter and relativistic Doppler. Consequently, our implementation uses an \textit{Extended Dynamic Mode Decomposition} (EDMD) Koopman estimator as a public-only filtering stage. Concretely, we form a delay-coordinate lifting of the beacon phase measurement $z_k$ into an embedding vector of dimension $d_K=16$ and fit a ridge-regularized linear propagator over a rolling window. This embedding dimension $d_K=16$ is selected to isolate dominant platform vibration modes and orbital harmonics while avoiding overfitting to optical shot-noise.

This public-only filtering yields a residual phase sequence $\phi_{\mathrm{res}}$ used by both the sheaf consistency score and the TDA pipeline. The phase portrait visually exposes the underlying dynamic attractor of the ISL telemetry, distinguishing structural degradation from temporary stochastic noise artifacts.

\begin{figure}[htbp]
    \centering
      \includegraphics[width=0.45\linewidth]{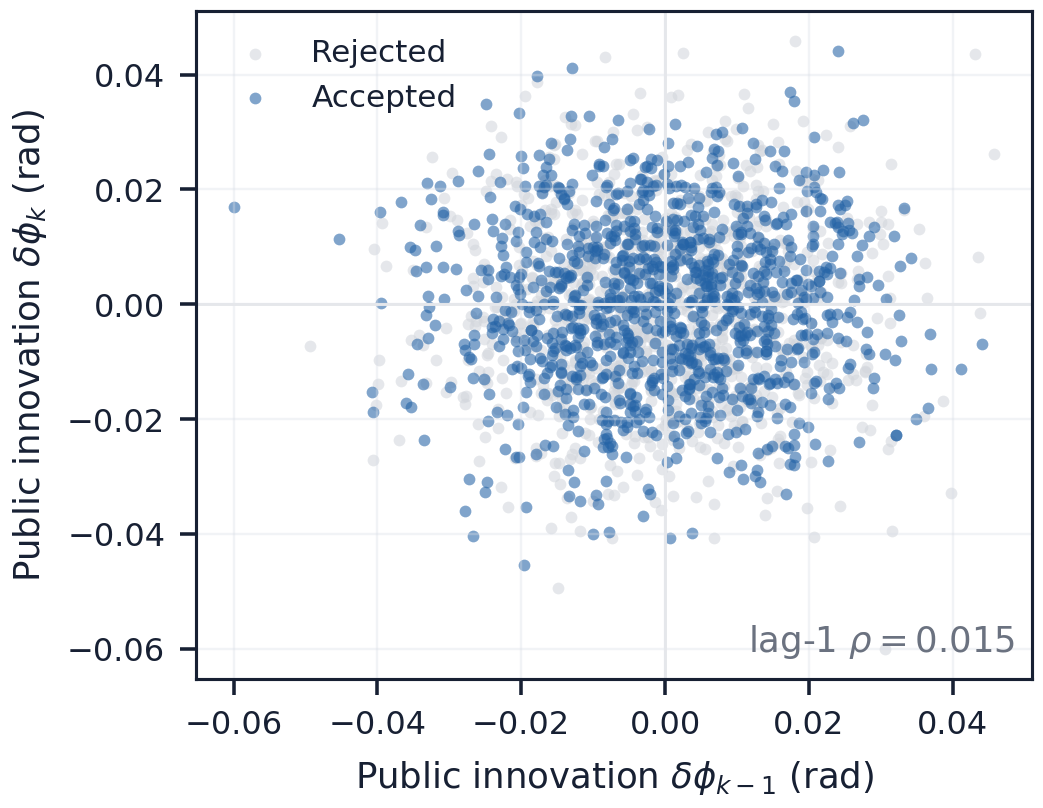}%
    \hfill
      \includegraphics[width=0.45\linewidth]{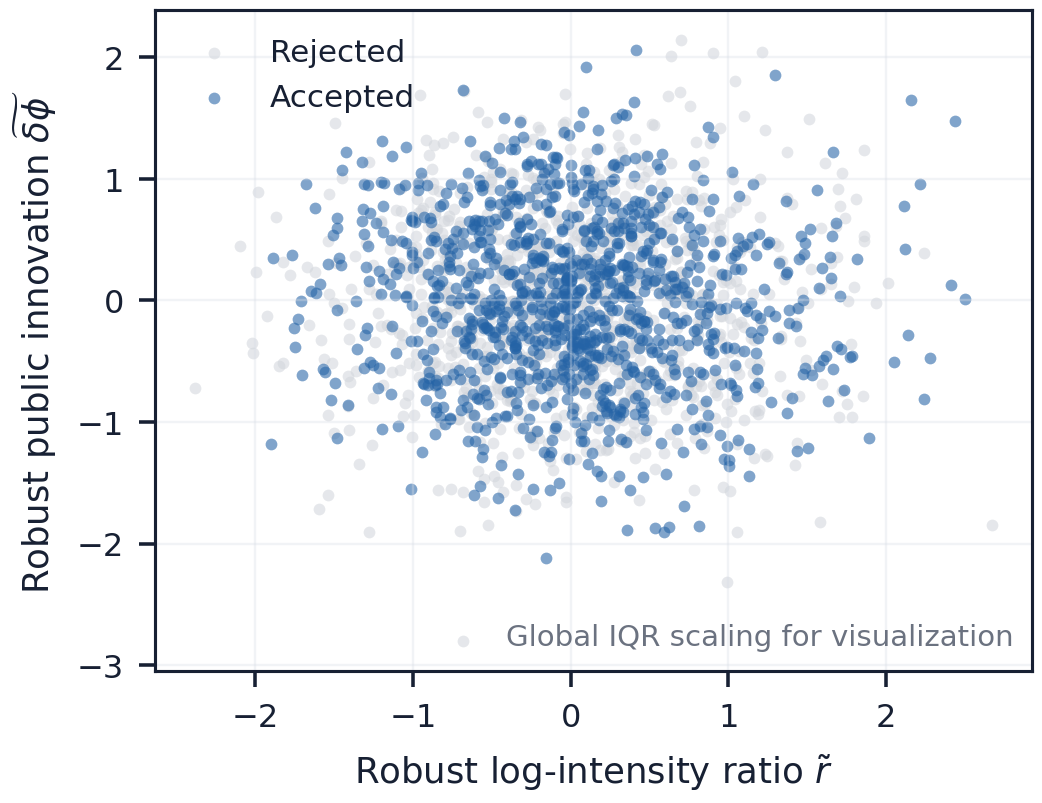}%    
    \caption{(Left) Public-only phase portrait representing the dynamic state space of the telemetry estimator. (Right) Robustly normalized telemetry point cloud used to build Alpha complexes for persistent homology.}
    \label{fig:manifolds}
\end{figure}

  \paragraph{Robust point cloud construction (IQR scaling)}
  Over a physical coherence window of $W$ blocks, we construct a 2D telemetry point cloud from \emph{reference-only} data:
  \begin{equation}
    r_k=\log\frac{I_{\mathrm{ref},A}(k)+\epsilon}{I_{\mathrm{ref},B}(k)+\epsilon},\qquad \phi_k=\phi_{\mathrm{res}}(k),
  \end{equation}
  then robustly normalize via interquartile ranges (IQR): $\tilde r_k=(r_k-\mathrm{med}(r))/\mathrm{IQR}(r)$ and $\tilde\phi_k=(\phi_k-\mathrm{med}(\phi))/\mathrm{IQR}(\phi)$. The point cloud is $\mathcal{X}=\{(\tilde r_k,\tilde\phi_k)\}_{k=1}^{W}$. Standardizing the raw telemetry via IQR scaling ensures that the topological features remain invariant to absolute distance attenuations, mapping diverse optical hardware outputs into a uniform, dimensionless metric space suitable for homology computation.

  \paragraph{Alpha complexes and persistence landscapes}
  Once $\mathcal{X}$ is constructed, we compute persistent homology using Alpha complexes (via Delaunay triangulations) \cite{Gudhi2014}. We focus on $H_1$ (cycles) since loop formation is the relevant signature of phase-inconsistent drift rather than mere connectedness. The persistence intervals are mapped into persistence landscapes \cite{Bubenik2015Landscapes}; we use the top three landscape layers and a fixed 100-bin resolution grid to obtain an $L_2$ energy
  \begin{equation}
    E_{\mathrm{TDA}}=\|\Lambda\|_2.
  \end{equation}
  Discretizing the landscape into a 100-bin grid ensures adequate filtration granularity for feature extraction, balancing resolution with the computational constraints of satellite-based processing.

  \paragraph{Density filtering and scalar TDA score}
  To suppress outliers and sparse point clouds, we compute a density proxy using a $k$-nearest-neighbor distance-to-measure (DTM) statistic (with $k=15$) and define a bounded scalar score
  \begin{equation}
    q_{\mathrm{TDA}}=\frac{E_{\mathrm{TDA}}}{\bigl(1+\mathbb{E}[\mathrm{DTM}]\bigr)\,\bigl(1+\mathbb{E}[\| (\tilde r,\tilde\phi) \|_2]\bigr)}.
  \end{equation}
  In the implemented oracle, $q_{\mathrm{TDA}}$ is converted into a trust factor $t_{\mathrm{TDA}}=\exp(-\kappa\,q_{\mathrm{TDA}})$ with $\kappa=5$, and this multiplicatively modulates the final public gate score. Mapping the landscape energy to a single bounded score streamlines the topological data, enabling the oracle to modulate block acceptance continuously rather than relying on binary thresholding. 

\begin{figure}[htbp]
    \centering
      \includegraphics[width=0.9\linewidth]{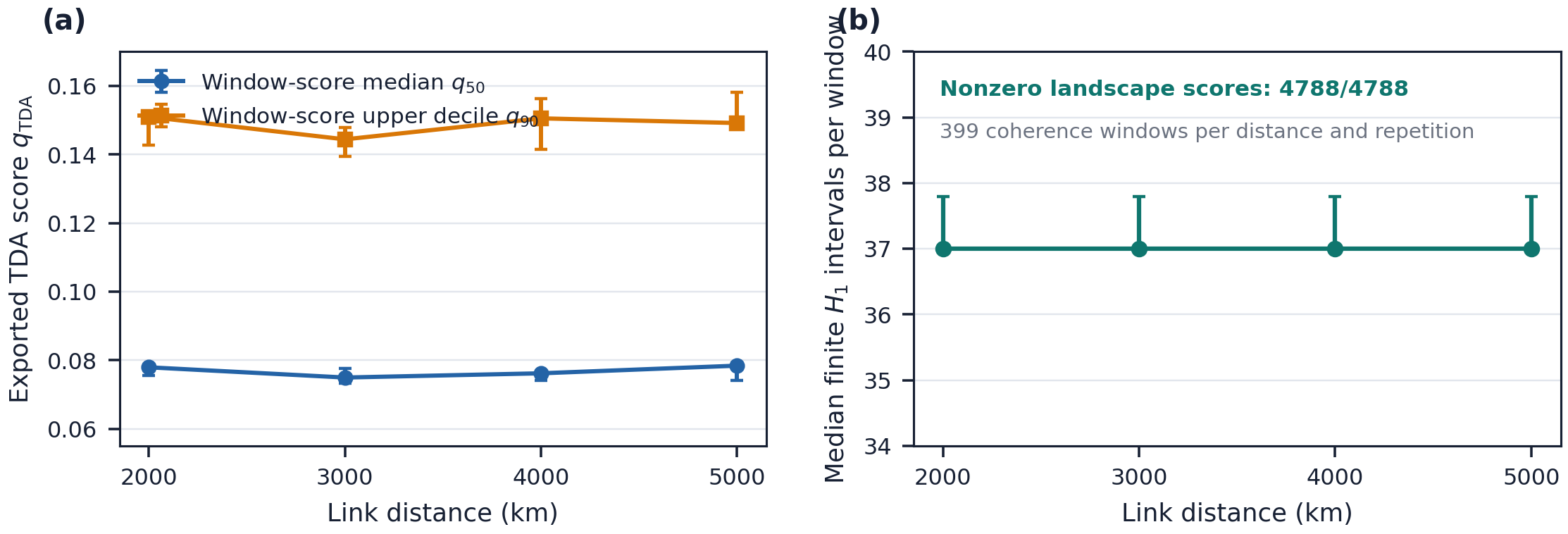}%
    \caption{Exported TDA diagnostics across the four distances. All 4,788 evaluated coherence windows produced nonzero persistence-landscape scores.}
    \label{fig:landscape}
\end{figure}

\subsection{Sheaf Hodge Decomposition as a Decentralized State Oracle}
\label{subsec:hodge_oracle}
Relying on hard cut-offs for individual edges introduces dangerous adaptive biases. Instead, we aggregate multiple public-only phase witnesses into a single sheaf-consistency trust score. In the minimal \emph{diamond} sub-complex (direct link $A \to B$ plus two witness paths $A \to C \to B$ and $A \to D \to B$), let $\phi_{\mathrm{dir}}$ be the public phase estimate for $A \to B$ and $\phi_{C},\phi_{D}$ the public witness phase proxies for the two alternate paths. We form residual 1-forms on $S^1$ via wrapped differences:
\begin{equation}
\begin{split}
    r_C &= \mathrm{wrap}(\phi_{\mathrm{dir}}-\phi_C), \\
    r_D &= \mathrm{wrap}(\phi_{\mathrm{dir}}-\phi_D), \\
    r_{CD} &= \mathrm{wrap}(\phi_C-\phi_D).
\end{split}
\end{equation}

\paragraph{Legacy obstruction norm (implemented baseline)}
The simplest global obstruction is a weighted edge-space norm
\begin{equation}
  e=\sqrt{r_C^2+r_D^2+\tfrac{1}{2}r_{CD}^2},
\end{equation}
mapped into a bounded trust score by a Lorentzian kernel with tolerance parameter $\tau$ (in radians):
\begin{equation}
  t_{\mathrm{sheaf}}=\frac{1}{1+(e/\tau)^2}.
\end{equation}
This design is deliberately conservative: large wrapped residuals cannot be ``canceled'' by signed averaging and therefore sharply reduce trust. While rudimentary, this unseparated norm provides a computationally lightweight fallback for highly symmetric topologies where the overhead of a full algebraic decomposition is unwarranted.

\paragraph{Hodge-mode cycle vs. gradient separation}
For stronger interpretability, the simulator optionally performs a closed-form discrete Hodge decomposition on the triangle edge-space \cite{JiangLimYaoYe2008HodgeRanking}. Using the cycle basis $c=[1,-1,1]$, the harmonic (cycle) coordinate is
\begin{equation}
  \gamma=\langle (r_C,r_D,r_{CD}),c\rangle/\langle c,c\rangle=(r_C-r_D+r_{CD})/3.
\end{equation}
The corresponding cycle energy and gradient energy are
\begin{equation}
\begin{split}
  E_{\mathrm{cyc}}&=\sqrt{3}\,|\gamma|,\\
  E_{\mathrm{grad}}&=\sqrt{(r_C-\gamma)^2+(r_D+\gamma)^2+(r_{CD}-\gamma)^2}.
\end{split}
\end{equation}
We penalize them independently with Lorentzian kernels (with potentially different tolerances) and set the global trust as a product
\begin{equation}
  t_{\mathrm{sheaf}}=\frac{1}{1+(E_{\mathrm{cyc}}/\tau_{\mathrm{cyc}})^2}\cdot\frac{1}{1+(E_{\mathrm{grad}}/\tau_{\mathrm{grad}})^2}.
\end{equation}
The scalar $E_{\mathrm{cyc}}$ is exported as a public-only \emph{cycle obstruction} diagnostic and can be used to define GEAT side-information states (Sec.~\ref{sec:sim_architecture}). Isolating the cycle obstruction ensures that the oracle measures global synchronization discrepancies independently of local phase fluctuations, translating these structural deviations into the discrete side-information states processed by the finite-key ledger.

\subsection{The Public Event $\Omega$ (Blind Thresholding)}
\label{subsec:public_event}
In our approach, the oracle defines a public acceptance event $\Omega$ for each temporal block interval based exclusively on public telemetry-derived scores. Operationally, the implemented gate score is the product
\begin{equation}
  g_i = t_{\phi,i}\,t_{\mathrm{sheaf},i}\,t_{\mathrm{TDA},i},
\end{equation}
where $t_{\phi,i}=\exp\bigl(- (\phi_{\mathrm{res},i}/0.15)^2\bigr)$ is a phase-consistency factor, $t_{\mathrm{sheaf},i}$ is the sheaf consensus score derived from the Hodge-Laplacian consistency witness, and $t_{\mathrm{TDA},i}=\exp(-5\,q_{\mathrm{TDA},i})$ is the persistence-landscape trust factor.

To avoid hand-tuning a brittle global threshold, the default simulator instantiation derives the acceptance threshold $\theta$ via a blind two-cluster KMeans split on the empirical gate score samples $\{g_i\}$, then accepts the high-score cluster:
\begin{equation}
  \Omega_i = \mathbb{I}\{ g_i \ge \theta \}.
\end{equation}

Crucially, because $I_{\mathrm{ref},A}, I_{\mathrm{ref},B}$ and the telemetry-derived phase estimator are public and independent from private key-basis operations, the event $\Omega$ satisfies the simulator contract \texttt{reference\_only\_gating}. It guarantees that the decision to harvest or discard a quantum transmission block uses zero key-basis click data. This separation forms the bedrock of our composable security claim: the network topology controls the sampling pool without creating adaptive leakage channels. How this uncompromised sampling pool is subsequently quantified into a strictly bounded cryptographic rate is the function of the Security Ledger Layer, detailed in the following section.

\section{Simulation Framework Architecture}
\label{sec:sim_architecture}

To bridge the gap between abstract quantum protocols and practical space deployment, we architect a modular simulation framework capable of seamlessly enforcing strict composable bounds under dynamic network conditions. The main deficit in existing TF-QKD simulators is their monolithic nature, deeply coupling static channel loss with asymptotic approximations. Our framework deliberately isolates the orbital substrate, the quantum execution, the topological control, and the security ledger layers. This guarantees that topological gating decisions ($\Omega$) are strictly audited as public transcripts before security generation. In this section, we detail the interplay of these four independent layers, from the physical geometry generation (Subsec. \ref{subsec:orbital_layer}) to the final composable ledger execution (Subsec. \ref{subsec:security_ledger}), concluding with an outline of our open-source artifact structure (Subsec. \ref{subsec:reproducibility}).

\subsection{Orbital Substrate Layer}
\label{subsec:orbital_layer}
The physical environment instantiation is governed by the \texttt{Orbital Substrate Layer}. Utilizing Two-Line Element (TLE) datasets, it propagates satellite trajectories to establish the time-dependent dynamic graph $G(V, E, t)$. This layer solves the space link budget on a discrete time-step basis, computing line-of-sight geometry, diffraction-induced free-space loss ($\eta_{\text{diff}}$), and stochastic pointing jitter probabilities ($\eta_{\text{ptr}}$). The continuous trajectory forms the exact transmittance distribution fed to the higher protocol layers.

\subsection{Quantum Execution Layer}
\label{subsec:quantum_layer}
The \texttt{Quantum Execution Layer} operates over the physical substrate, modeling the execution of the TF-QKD protocol across the fluctuating links. Incorporating the standard decoy-state apparatus (signal, decoy, and vacuum intensities) \cite{LoMaChen05}, it simulates the distribution of coherent states and computes realistic interference statistics at the central node (a satellite relay or ground station). By ingesting the instantaneous parameters derived from the Orbital Substrate Layer, it handles practical detector characteristics, probabilistic dark counts, and visibility degradation caused by Doppler shift transients. 

\paragraph{Finite-Key Stochastic Noise Injection}
A critical structural divergence in our framework from standard simulators is the explicit mathematical rejection of asymptotic probabilities applied linearly to finite blocks. To guarantee cryptographic rigor, our quantum layer implements a strict stochastic discretization routine (\texttt{stochasticize\_block\_metrics}). For any given $N_{\text{block}}$, the raw expected yields (e.g., $Q_{\mu}, Q_{\nu_1}$) and phase-error probabilities ($e_{\text{ph}}$) are not utilized as ground truth. Instead, they form the expectation parameters for Binomial sampling distributions ($\mathcal{B}(n, p)$). 

For example, the Z-basis intensity trial sizes are sampled explicitly as $N_{\mu} = N_{\text{block}} \cdot p_z^2 \cdot P(\mu|\text{Z})$. The observed clicks are then generated via binomial throws $c \sim \mathcal{B}(N_{\mu}, Q_{\mu})$, converting theoretical probabilities into noisy, discrete empirical frequencies $\tilde{Q}_{\mu} = c/N_{\mu}$. Similarly, X-basis errors are sampled conditionally on the stochastically derived X-basis click rates. This ensures the generalized entropy (GEAT) min-tradeoff functions are fed realistic, highly turbulent finite-key observables rather than artificially smooth averages \cite{Kamin2024GEATDecoy}. Consequently, omitting this stochastic noise injection leads to severe underestimations of finite-size penalties, compromising the validity of non-IID satellite QKD simulations.

\subsection{Topological Control Layer}
\label{subsec:control_layer}
Unlike conventional simulators that pipeline quantum click events directly into identical post-processing routines, our architecture structurally isolates this flow. The \texttt{Topological Control Layer} acts as the decentralized State Oracle. It constructs the Cellular Sheaf detailed in Section~\ref{sec:topological_protocol} using purely public reference signals (beacon/reference phases and reference intensities). For each block $i$, the oracle computes a public gate score $g_i=t_{\phi,i}\,t_{\mathrm{sheaf},i}\,t_{\mathrm{TDA},i}$: a phase-consistency trust factor from the public estimator residual, a sheaf-consensus trust factor from the diamond witness paths (optionally decomposed into cycle vs gradient obstruction), and a persistence-landscape trust factor computed on a coherence-window point cloud derived from $(\log(I_{\mathrm{ref},A}/I_{\mathrm{ref},B}),\,\phi_{\mathrm{res}})$.

The acceptance mask $\Omega$ is then derived either by a blind two-cluster KMeans thresholding of the empirical samples $\{g_i\}$ (default), or by enforcing a target acceptance fraction (top-$k$ by score). By design, the oracle never accesses private key-basis click data, averting information leakage side-channels and preserving the Markov structure required by the GEAT.

\subsection{Security Ledger Layer}
\label{subsec:security_ledger}
Evaluating dynamic space networks requires confronting finite-size statistics in non-IID conditions. Our framework mandates a \texttt{Security Ledger Layer} explicitly built on the Entropy Accumulation paradigms \cite{ArnonFriedman18EAT, MetgerRenner2022QKDGEA}. Conceptually, this layer operates analogously to a strict cryptographic accounting system. Every block of pulses collected represents potential raw key material; however, statistical fluctuations, error-correction leakage, and composable security limits act as heavy operational taxes. Instead of artificially aggregating all transmissions and uniformly applying independent and identically distributed (IID) laws—a physically flawed assumption under dynamic orbital turbulence—the ledger explicitly accounts for these shifting conditions, ensuring the available secure entropy is never mathematically over-drafted.

\paragraph{The Markov Confidence Condition}
To claim composable security under topological gating, the post-selection must not leak information about the raw key bases. This requires a strict Markov condition: the public telemetry $S_i$ (and its derived gating decision $T_i \in \{0,1\}$) must be conditionally independent of the private operations $X_i, Z_i$, given the quantum state $\rho_i$. Our framework enforces this orthogonality structurally: the Topological Control Layer operates exclusively on auxiliary bright pulses and classical beacon tracking, shielding the quantum click data from the acceptance logic.

\paragraph{Algorithmic Ledger Derivation and LP Solver}
The Security Ledger Layer segments the protocol into $m$ operational blocks. For blocks surviving the $\Omega=1$ gating operation, it computes the secure key length $\ell$ bound by the GEAT framework:
\begin{equation}
    \ell \ge \sum_{i=1}^n \min f(p_{\text{obs},i}) - \Delta_{\text{GEAT}} - \mathrm{leak}_{\text{EC}} - \log_2(1/\varepsilon)
\end{equation}
where $f(\cdot)$ is the min-tradeoff function evaluated over the observed error rates $p_{\text{obs},i}$. Crucially, we abandon asymptotic approximations in favor of strict finite-size confidence intervals. Applying Hoeffding, Clopper-Pearson, and Serfling bounds \cite{ClopperPearson34, Hoeffding63, Serfling1974} specifically adapted for non-IID conditional blocks, we rigorously bound the statistical fluctuation of phase errors. By processing only the $n$ signals surviving the gating operation, the ledger constructs a worst-case parameter regime.

To execute the $\min_{p} f(p)$ bound practically, our architecture leverages a Numerical Linear Programming (LP) continuous engine, specifically crafted to optimize the transcript symbols under Bernstein bounds. The solver enforces strict violation constraints ($10^{-8}$ numeric tolerance) to guarantee no floating-point arithmetic flaws introduce security vulnerabilities into the ledger. This numeric precision is vital: standard convex optimizers often fail to converge properly when processing extremely skewed non-IID parameters in LEO regimes. Our strict LP bound ensures that any resulting positive output rate represents a structurally sound cryptographic proof, robust against arithmetic drift and eliminating the over-optimistic rate estimations inherent in traditional QKD architectures.

\subsection{Artifact Availability and Reproducibility}
\label{subsec:reproducibility}
To make the claims in this paper directly auditable, we publish a complete artifact bundle via Zenodo: \ZenodoDOI. The bundle includes: (i) the reference implementation entrypoint (\texttt{eat\_isl\_sim.py}), (ii) the exact configuration and command-line arguments used to generate all paper figures, and (iii) the raw numeric exports (JSON) from which the plots are produced.

\paragraph{Provenance and Determinism}
Every benchmark run exports a machine-readable provenance record (\texttt{isl\_data.meta.json}) containing the full argument vector, the git commit hash, the complete parameter object, and the exact Python package versions. The corresponding numeric output (\texttt{isl\_data.json}) stores the per-distance SKR values, acceptance rates, link-budget decompositions, and (when enabled) compact diagnostics and time-series traces. This design ensures that every plotted point is reproducible from artifacts alone, without undocumented manual intervention.

\paragraph{Reproduction Recipe (Single-Command)}
Reproducing the primary ISL figure is achieved by re-executing the recorded argument vector embedded in \texttt{isl\_data.meta.json}. Concretely, the run used for Figure~\ref{fig:eat_isl} is an ISL, block-HMM, composable finite-key execution with explicit count sampling (\texttt{--simulate-counts}) and a transcript-driven GEAT engine (\texttt{--eat-engine numerical\_geat}) that constructs a finite transcript alphabet and derives fixed per-symbol min-tradeoff bounds via a local decoy LP (\texttt{--geat-tradeoff-mode transcript\_symbol\_lp\_lower\_bernstein}).

This paper therefore separates \emph{scientific claims} (protocol and security modeling) from \emph{reproducibility claims} (artifact completeness): the former is argued in Sections~\ref{sec:topological_protocol} and \ref{sec:evaluation}, while the latter is guaranteed by exported provenance and immutable artifact snapshots.

\section{Interpretation and Discussion}
\label{sec:evaluation}

We deploy the simulation framework in a dynamic Low Earth Orbit (LEO) inter-satellite-link setting with two orbital transmitters ($A$, $B$) and a central, untrusted space relay ($C$). The same stochastic model is evaluated at fixed separations of 2,000, 3,000, 4,000, and 5,000 km. In this section, we detail the experimental configuration (Subsec. \ref{subsec:exp_config}), define the TDA ablation (Subsec. \ref{subsec:baselines}), present the finite-size rate extraction results (Subsec. \ref{subsec:results_skr}), and discuss their interpretation and limitations (Subsecs. \ref{subsec:interpretations} and \ref{subsec:tradeoffs}).

\subsection{Experimental Configuration}
\label{subsec:exp_config}
The physical layer parameters intentionally expose dynamic, non-IID conditions:
\begin{itemize}
    \item We define transmission and reception aperture diameters of 0.40 m and operating wavelengths at standard telecom bounds (1550 nm), mapping exactly to high-efficiency deep-space optical terminals.
    \item Pointing jitter induces stochastic angular misalignments, modeled as a continuous Gaussian process scaled to a highly constrained 0.3 $\mu$rad. 
    \item Super-conducting detectors operate with a baseline system efficiency ($\eta_{\text{sys}} = 0.5$) alongside ultra-low dark count rates ($p_d = 10^{-11}$ per pulse).
    \item A system repetition rate of 1.0 GHz dictates an aggregated evaluation block of $N = 10^{13}$ protocol pulses partitioned into $2 \times 10^4$ discrete segments along the orbital pass.
\end{itemize}

To rigorously reproduce the evaluation scenario, we outline the exact physical constraints mapping the evaluation instance in Table~\ref{tab:sim_parameters}. Note that the baseline performance metrics depend strictly on these initial state inputs, scaling as a function of continuous geometric degradation via orbital altitude and aperture bounds.

\begin{table}[htbp]
\renewcommand{\arraystretch}{1.2}
\caption{ISL Simulation Constraints and Parameters}
\label{tab:sim_parameters}
\centering
\begin{tabular}{ll|ll}
\toprule
\textbf{Parameter} & \textbf{Value} & \textbf{Parameter} & \textbf{Value} \\ 
\midrule
Repetition Rate ($\nu$) & 1.0 GHz & Tot. Pulses ($N$) & $10^{13}$ \\
System Eff. ($\eta_{\text{sys}}$) & 0.50 & Blocks & $2\times 10^4$ \\
Dark Count ($p_{d}$) & $10^{-11}$ & Sec. Margin ($\varepsilon$) & $10^{-10}$ \\ 
Aperture TX/RX & 0.40 m & Atmos. Loss & 0.0 dB \\
Wavelength ($\lambda$) & 1550 nm & Error Corr. & $f_{ec} = 1.16$ \\
Pointing Jitter & 0.3 $\mu$rad & TDA Window & 50 blocks \\
Phase Offset & $\mathcal{N}(0, \sigma_{\phi}^2)$ & TDA Stride & 50 blocks \\
\bottomrule
\end{tabular}
\end{table}

These parameters were meticulously selected to guide the reader through a realistic, rather than overly optimistic, deep-space quantum deployment. For instance, the 0.40 m apertures and 0.3 $\mu$rad pointing jitter represent the state-of-the-art in precision satellite tracking, essential for maintaining Single Mode Fiber (SMF) coupling. The block division ($2 \times 10^4$) is not arbitrary; it guarantees that the tracking telemetry provides a continuous, high-resolution dataset to feed into the Topological Control Layer, allowing the system to safely gate the incoming transmission without violating GEAT limits. Each distance is evaluated with three Monte Carlo repetitions, explicit count sampling, Koopman embedding dimension 16, Hodge mode, and the full transcript-symbol LP lower bound.

\subsection{Comparative Baseline}
\label{subsec:baselines}
To isolate the effect of persistent homology, we compare two configurations over identical physical channel realizations and ledger settings:
\begin{enumerate}
    \item \textbf{TDA off:} the public gate retains the causal Koopman EDMD phase estimator and rank-two Hodge cycle score, while the persistence-landscape factor is explicitly disabled.
    \item \textbf{TDA on:} the same public gate additionally includes the persistence-landscape trust factor. Across all repetitions, 4,788 of 4,788 evaluated TDA windows produce nonzero scores.
\end{enumerate}

\subsection{Results: Finite-Size Rate Extraction}
\label{subsec:results_skr}
\begin{figure*}[t]
    \centering
      \includegraphics[width=0.94\textwidth]{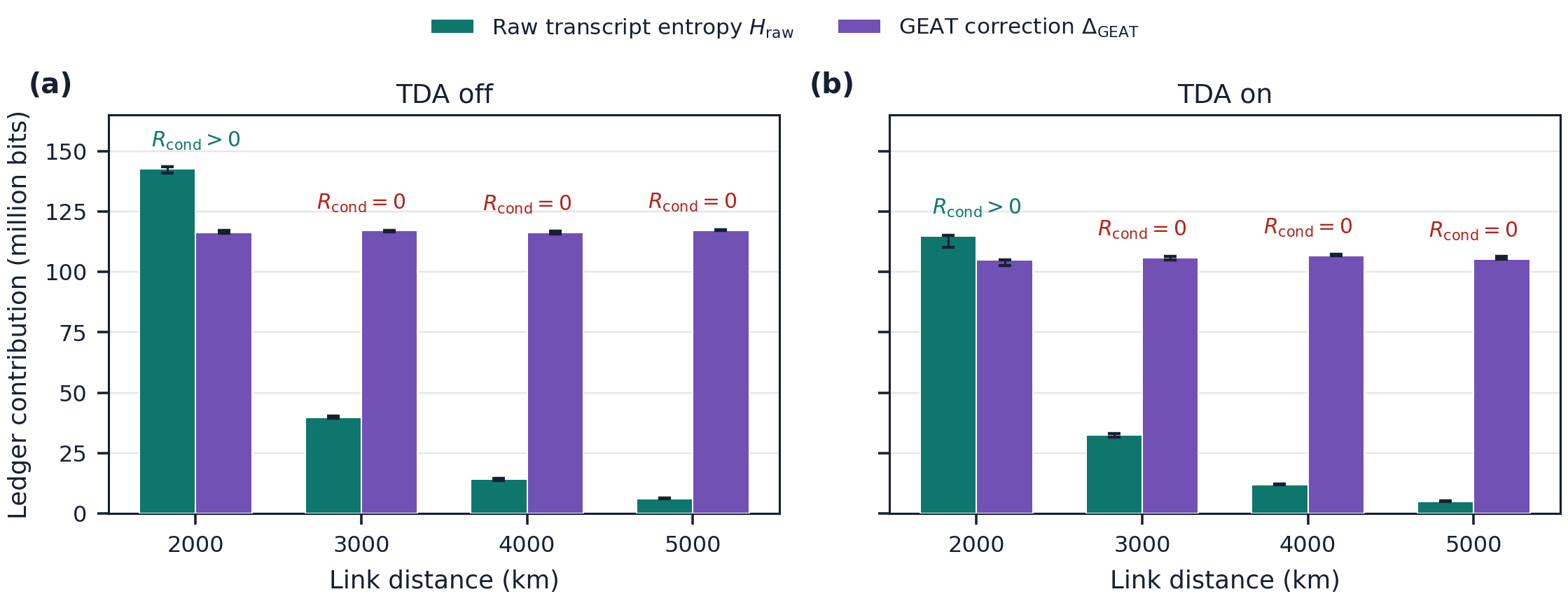}%
    \caption{Raw transcript-entropy lower bound and GEAT correction for the TDA-off and TDA-on configurations. Bars show medians and whiskers show the Monte Carlo q10--q90 interval over three repetitions.}
    \label{fig:results_skr}
\end{figure*}

Figure~\ref{fig:results_skr} exposes the finite-size ledger directly. At 2,000 km the raw transcript-entropy lower bound exceeds the GEAT correction in both configurations. From 3,000 km onward, the GEAT correction exceeds the raw contribution before error-correction leakage and privacy-amplification terms are subtracted, and the conditional candidate is therefore clipped to zero.

\begin{figure*}[t]
    \centering
      \includegraphics[width=0.94\textwidth]{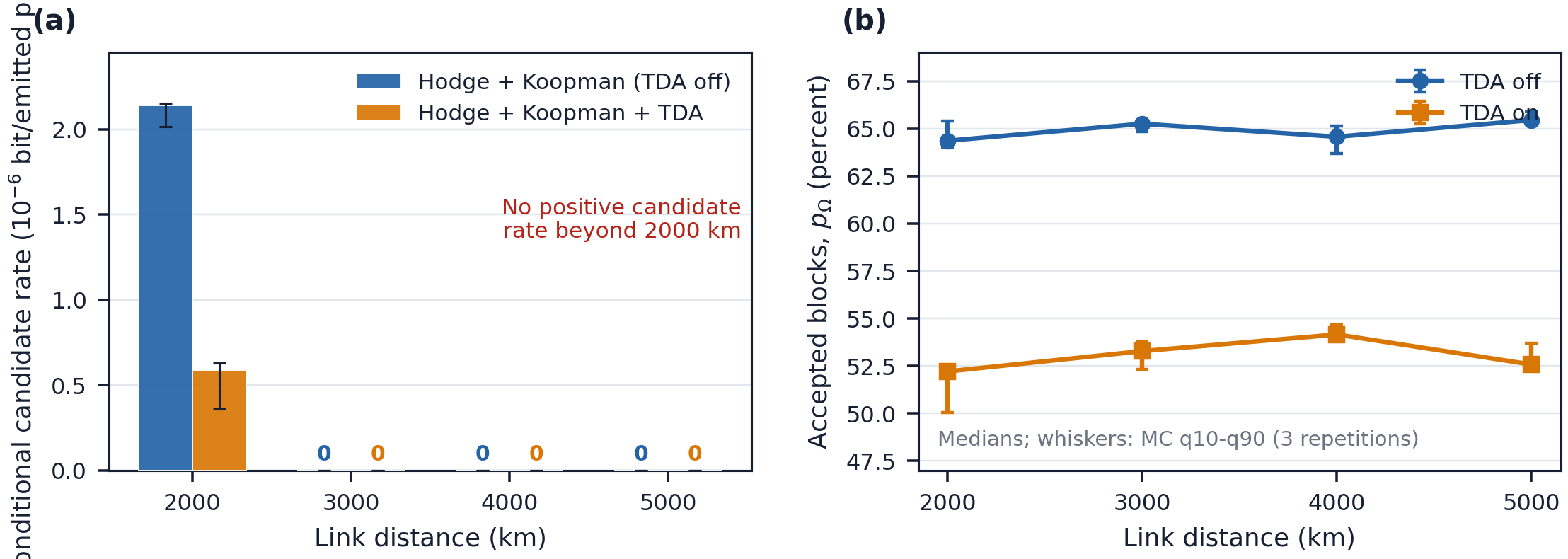}%
    \caption{Conditional candidate rate and accepted-block fraction for the same Hodge--Koopman gate with TDA disabled or enabled. Bars and points show medians; whiskers show q10--q90 over three Monte Carlo repetitions.}
    \label{fig:eat_isl}
\end{figure*}

Figure~\ref{fig:eat_isl} reports the resulting rates and acceptance fractions. The median conditional candidate at 2,000 km is $2.14\times10^{-6}$ bit per emitted pulse with TDA off and $5.87\times10^{-7}$ with TDA on. Both configurations return zero at 3,000, 4,000, and 5,000 km. At 2,000 km, the corresponding median accepted fractions are 64.36\% and 52.22\%, respectively.

The formal-security status exported by every sample is \texttt{CONDITIONAL\_PROTOCOL\_PROOF\_INCOMPLETE}. Accordingly, the positive 2,000-km values are conditional numerical candidates; the certified composable rate is zero at every evaluated distance.

\subsection{Physical and Cryptographic Interpretations}
\label{subsec:interpretations}
The ablation does not show a rate advantage from TDA under the selected parameters. Enabling it lowers the 2,000-km median candidate by 72.6\% and reduces the median acceptance fraction from 64.36\% to 52.22\%; it does not extend the positive-candidate range. This is a negative but informative result: persistent homology is numerically active, yet its additional selectivity is not compensated by a sufficiently improved accepted transcript in this scenario \cite{DupuisFawziRenner16EAT}.

The zeros at 3,000--5,000 km should not be read as a general impossibility result for TF-QKD or for ISLs. They apply to the specified apertures, pulse budget, stochastic channel, public gate, and finite-size ledger. The exact LP audit reports no \texttt{infeasible\_zero} contribution in any of the 24 samples, so these zeros are not caused by conservative masking of infeasible LP boxes.

\subsection{Systemic Trade-offs and Architectural Limitations}
\label{subsec:tradeoffs}
The TDA path constructs Alpha complexes on two-dimensional telemetry windows after the causal 16-dimensional Koopman embedding has produced the public phase innovation. This adds processing and reduces acceptance, so an operational design would require calibration against a declared objective rather than treating TDA as intrinsically beneficial.

The present comparison also isolates only the TDA factor: Hodge and Koopman remain active in both arms. It therefore supports no claim about superiority over conventional thresholding, nor does it establish that rejected blocks were physically insecure.

The result instead provides an auditable benchmark for improving the gate: future work must demonstrate an acceptance-matched or security-relevant advantage before persistent homology can be credited with an operational range benefit.

\section{Conclusions and Future Work}
\label{sec:conclusions}

Scaling Twin-Field QKD from laboratory demonstrations to dynamic space-based quantum networks creates a profound methodological intersection between moving network topologies and finite-key cryptographic bounds. Attempting to simulate such deployments while treating topological post-selection ($\Omega$) as an implicit, loosely coupled heuristic introduces critical composability flaws. In this paper, we bridge this structural gap by presenting a comprehensive simulation framework that implements a novel topological protocol based on Cellular Sheaves.

By equipping the dynamic satellite constellation with a distributed sheaf, a non-linear Koopman embedding, and computing the discrete Hodge Laplacian, we obtain a rigorously deterministic, algebraically sound Topological Control Layer. This mechanism operates entirely on public telemetry, preventing the systemic information leakages typical of adaptive state-filtering architectures. Evaluating this method through our strictly partitioned Security Ledger Layer shows positive conditional numerical candidates only at 2,000 km: the median is $2.14\times10^{-6}$ bit per emitted pulse with TDA disabled and $5.87\times10^{-7}$ with TDA enabled, while both configurations return zero at 3,000--5,000 km. In this scenario, TDA reduces the candidate rate and accepted fraction and does not extend the positive-candidate range. The full protocol-level composable-security theorem remains incomplete, so the certified composable rate is zero throughout. 

These results underscore a crucial engineering requirement for global quantum networks: algorithmic topological administration must be evaluated jointly with its acceptance loss and finite-size ledger, rather than assumed to provide an intrinsic rate advantage. The present TDA ablation is therefore an auditable negative result and a baseline for future acceptance-matched optimization. 

\subsection{Future Trajectories}
Building upon this state-aware framework, significant research trajectories emerge in Quantum Software Engineering (QSE) and space operations:
\begin{itemize}
    \item \textbf{Formal Verification of Oracle Models:} The strict conditional independence dictated by the Markov confidence limits demands machine-verifiable proofs (e.g., via Lean 4 or Coq), guaranteeing mathematically zero key-leakage through the public acceptance gate $\Omega$. 
    \item \textbf{Distributed Hardware Acceleration:} Generating homological point clouds via Delaunay triangulations and Alpha complexes requires significant continuous processing. Accelerating the TDA pipeline (and GUDHI primitives) on radiation-hardened FPGA architectures within the satellite bus will be essential to reduce end-to-end processing delays in orbit.
    \item \textbf{Dynamic Routing in Mega-Constellations:} By continuously mapping the ISL state to a Hodge Laplacian, future protocols can optimize quantum routing across large-scale satellite swarms, actively diverting entanglement generation away from structural topological failures.
\end{itemize}

\bibliographystyle{IEEEtran}
\bibliography{references} 

\begin{thebibliography}{10}
\providecommand{\url}[1]{#1}
\csname url@samestyle\endcsname
\providecommand{\newblock}{\relax}
\providecommand{\bibinfo}[2]{#2}
\providecommand{\BIBentrySTDinterwordspacing}{\spaceskip=0pt\relax}
\providecommand{\BIBentryALTinterwordstretchfactor}{4}
\providecommand{\BIBentryALTinterwordspacing}{\spaceskip=\fontdimen2\font plus
\BIBentryALTinterwordstretchfactor\fontdimen3\font minus
  \fontdimen4\font\relax}
\providecommand{\BIBforeignlanguage}[2]{{%
\expandafter\ifx\csname l@#1\endcsname\relax
\typeout{** WARNING: IEEEtran.bst: No hyphenation pattern has been}%
\typeout{** loaded for the language `#1'. Using the pattern for}%
\typeout{** the default language instead.}%
\else
\language=\csname l@#1\endcsname
\fi
#2}}
\providecommand{\BIBdecl}{\relax}
\BIBdecl

\bibitem{Liao2017Satellite}
S.-K. Liao and et~al., ``Satellite-to-ground quantum key distribution,''
  \emph{Nature}, vol. 549, pp. 43--47, 2017.

\bibitem{Yin2020Satellite}
J.~Yin and et~al., ``Entanglement-based secure quantum cryptography over 1,120
  kilometres,'' \emph{Nature}, vol. 582, pp. 501--505, 2020.

\bibitem{Pirandola2021FreeSpaceLimits}
\BIBentryALTinterwordspacing
S.~Pirandola, ``Limits and security of free-space quantum communications,''
  \emph{Physical Review Research}, vol.~3, p. 013279, 2021. [Online].
  Available: \url{https://arxiv.org/abs/2010.04168}
\BIBentrySTDinterwordspacing

\bibitem{Vergoossen2019SatConstellations}
\BIBentryALTinterwordspacing
T.~Vergoossen, S.~Loarte, R.~Bedington, H.~Kuiper, and A.~Ling, ``Satellite
  constellations for trusted node qkd networks,'' \emph{Acta Astronautica},
  2020. [Online]. Available: \url{https://arxiv.org/abs/1903.07845}
\BIBentrySTDinterwordspacing

\bibitem{PirandolaPLOB17}
S.~Pirandola, R.~Laurenza, C.~Ottaviani, and L.~Banchi, ``Fundamental limits of
  repeaterless quantum communications,'' \emph{Nature Communications}, vol.~8,
  p. 15043, 2017.

\bibitem{Lucamarini18}
M.~Lucamarini, Z.~Yuan, J.~F. Dynes, and A.~J. Shields, ``Overcoming the
  rate--distance limit of quantum key distribution without quantum repeaters,''
  \emph{Nature}, vol. 557, pp. 400--403, 2018.

\bibitem{MaSNS18}
X.-B. Wang, Z.-W. Yu, and X.-L. Hu, ``Twin-field quantum key distribution with
  large misalignment error,'' \emph{Physical Review A}, vol.~98, no.~6, p.
  062323, 2018.

\bibitem{Curty19}
M.~Curty, K.~Azuma, and H.-K. Lo, ``Simple security proof of twin-field quantum
  key distribution protocols,'' \emph{npj Quantum Information}, vol.~5, no.~1,
  p.~64, 2019.

\bibitem{Liu2023TF1000km}
\BIBentryALTinterwordspacing
Y.~Liu, W.-J. Zhang, C.~Jiang, J.-P. Chen, C.~Zhang, W.-X. Pan, D.~Ma, H.~Dong,
  J.-M. Xiong, C.-J. Zhang, H.~Li, R.-C. Wang, J.~Wu, T.-Y. Chen, L.~You, X.-B.
  Wang, Q.~Zhang, and J.-W. Pan, ``Experimental twin-field quantum key
  distribution over 1000 km fiber distance,'' \emph{Physical Review Letters},
  vol. 130, p. 210801, 2023. [Online]. Available:
  \url{https://arxiv.org/abs/2303.15795}
\BIBentrySTDinterwordspacing

\bibitem{Li2025FreeSpaceTFQKD}
\BIBentryALTinterwordspacing
Y.-H. Li, T.~Zeng, M.-Y. Wang, C.~Jiang, J.~Lin, H.-B. Fu, X.-Y. Zheng, J.-P.
  Chen, Z.-S. Lin, C.-L. Li, J.-Y. Guan, Y.~Li, Q.~Shen, H.~Li, L.~You,
  Z.~Wang, F.~Zhou, J.~Yin, S.-K. Liao, J.-G. Ren, X.-B. Wang, Y.~Cao,
  Q.~Zhang, C.-Z. Peng, and J.-W. Pan, ``Free-space twin-field quantum key
  distribution,'' arXiv:2503.17744, 2025. [Online]. Available:
  \url{https://arxiv.org/abs/2503.17744}
\BIBentrySTDinterwordspacing

\bibitem{Cao2020FreeSpaceMDIQKD}
\BIBentryALTinterwordspacing
Y.~Cao, Y.-H. Li, K.-X. Yang, Y.-F. Jiang, S.-L. Li, X.-L. Hu, M.~Abulizi,
  C.-L. Li, W.~Zhang, Q.-C. Sun, W.-Y. Liu, X.~Jiang, S.-K. Liao, J.-G. Ren,
  H.~Li, L.~You, Z.~Wang, J.~Yin, C.-Y. Lu, X.-B. Wang, Q.~Zhang, C.-Z. Peng,
  and J.-W. Pan, ``Long-distance free-space measurement-device-independent
  quantum key distribution,'' arXiv:2006.05088, 2020. [Online]. Available:
  \url{https://arxiv.org/abs/2006.05088}
\BIBentrySTDinterwordspacing

\bibitem{Shan2024SNSPhasePostselection}
\BIBentryALTinterwordspacing
Y.-G. Shan, Y.~Zhou, Z.-Q. Yin, and S.~Wang, ``Sending-or-not-sending quantum
  key distribution with phase postselection,'' arXiv:2401.02304, 2024.
  [Online]. Available: \url{https://arxiv.org/abs/2401.02304}
\BIBentrySTDinterwordspacing

\bibitem{MetgerRenner2022QKDGEA}
\BIBentryALTinterwordspacing
T.~Metger and R.~Renner, ``Security of quantum key distribution from
  generalised entropy accumulation,'' 2022, arXiv:2203.04993. [Online].
  Available: \url{https://arxiv.org/abs/2203.04993}
\BIBentrySTDinterwordspacing

\bibitem{Metger2022GEA}
\BIBentryALTinterwordspacing
T.~Metger, O.~Fawzi, D.~Sutter, and R.~Renner, ``Generalised entropy
  accumulation,'' 2022, arXiv:2203.04989. [Online]. Available:
  \url{https://arxiv.org/abs/2203.04989}
\BIBentrySTDinterwordspacing

\bibitem{HansenGhrist2019CellularSheaves}
\BIBentryALTinterwordspacing
J.~Hansen and R.~Ghrist, ``Toward a spectral theory of cellular sheaves,''
  2019. [Online]. Available: \url{https://arxiv.org/abs/1808.01513}
\BIBentrySTDinterwordspacing

\bibitem{Carlsson2009TDA}
G.~Carlsson, ``Topology and data,'' \emph{Bulletin of the American Mathematical
  Society}, vol.~46, no.~2, pp. 255--308, 2009.

\bibitem{Chen2021TF511km}
\BIBentryALTinterwordspacing
J.-P. Chen, C.~Zhang, Y.~Liu, C.~Jiang, W.-J. Zhang, Z.-Y. Han, S.-Z. Ma, X.-L.
  Hu, Y.-H. Li, H.~Liu, F.~Zhou, H.-F. Jiang, T.-Y. Chen, H.~Li, L.-X. You,
  Z.~Wang, X.-B. Wang, Q.~Zhang, and J.-W. Pan, ``Twin-field quantum key
  distribution over 511 km optical fiber linking two distant metropolitans,''
  \emph{Nature Photonics}, vol.~15, p. 570, 2021. [Online]. Available:
  \url{https://arxiv.org/abs/2102.00433}
\BIBentrySTDinterwordspacing

\bibitem{AndrewsPhillips2005}
L.~C. Andrews and R.~L. Phillips, \emph{Laser Beam Propagation through Random
  Media}, 2nd~ed.\hskip 1em plus 0.5em minus 0.4em\relax SPIE Press, 2005.

\bibitem{Rabiner1989HMM}
L.~R. Rabiner, ``A tutorial on hidden markov models and selected applications
  in speech recognition,'' \emph{Proceedings of the IEEE}, vol.~77, no.~2, pp.
  257--286, 1989.

\bibitem{Zhou2023PhaseErrorPostselection}
\BIBentryALTinterwordspacing
Y.~Zhou, Z.-Q. Yin, Y.-G. Shan, and Z.-H. Wang, ``Precise phase error rate
  analysis for quantum key distribution with phase postselection,''
  arXiv:2312.06385, 2023. [Online]. Available:
  \url{https://arxiv.org/abs/2312.06385}
\BIBentrySTDinterwordspacing

\bibitem{ArnonFriedman18EAT}
R.~Arnon-Friedman, F.~Dupuis, O.~Fawzi, R.~Renner, and T.~Vidick, ``Practical
  device-independent quantum cryptography via entropy accumulation,''
  \emph{Nature Communications}, vol.~9, no.~1, p. 459, 2018.

\bibitem{Kamin2024GEATDecoy}
\BIBentryALTinterwordspacing
L.~Kamin, A.~Arqand, I.~George, N.~L{"u}tkenhaus, and E.~Y.-Z. Tan,
  ``Finite-size analysis of prepare-and-measure and decoy-state qkd via entropy
  accumulation,'' arXiv:2406.10198, 2024. [Online]. Available:
  \url{https://arxiv.org/abs/2406.10198}
\BIBentrySTDinterwordspacing

\bibitem{Curry2014Sheaves}
J.~M. Curry, ``Sheaves, cosheaves and applications,'' 2014.

\bibitem{Zomorodian2005}
A.~J. Zomorodian, \emph{Topology for Computing}.\hskip 1em plus 0.5em minus
  0.4em\relax Cambridge University Press, 2005.

\bibitem{CohenSteiner2007}
D.~Cohen-Steiner, H.~Edelsbrunner, and J.~Harer, ``Stability of persistence
  diagrams,'' \emph{Discrete \& Computational Geometry}, vol.~37, no.~1, pp.
  103--120, 2007.

\bibitem{Gudhi2014}
C.~Maria, J.-D. Boissonnat, M.~Glisse, and M.~Yvinec, ``The {Gudhi} library:
  Simplicial complexes and persistent homology,'' in \emph{Mathematical
  Software -- {ICMS} 2014}, ser. Lecture Notes in Computer Science, vol.
  8592.\hskip 1em plus 0.5em minus 0.4em\relax Springer, 2014, pp. 167--174.

\bibitem{Bubenik2015Landscapes}
P.~Bubenik, ``Statistical topological data analysis using persistence
  landscapes,'' \emph{Journal of Machine Learning Research}, vol.~16, no.~1,
  pp. 77--102, 2015.

\bibitem{JiangLimYaoYe2008HodgeRanking}
\BIBentryALTinterwordspacing
X.~Jiang, L.-H. Lim, Y.~Yao, and Y.~Ye, ``Statistical ranking and combinatorial
  hodge theory,'' 2009. [Online]. Available:
  \url{https://arxiv.org/abs/0811.1067}
\BIBentrySTDinterwordspacing

\bibitem{LoMaChen05}
H.-K. Lo, X.~Ma, and K.~Chen, ``Decoy state quantum key distribution,''
  \emph{Phys. Rev. Lett.}, vol.~94, no.~23, p. 230504, 2005.

\bibitem{ClopperPearson34}
C.~J. Clopper and E.~S. Pearson, ``The use of confidence or fiducial limits
  illustrated in the case of the binomial,'' \emph{Biometrika}, vol.~26, no.~4,
  pp. 404--413, 1934.

\bibitem{Hoeffding63}
W.~Hoeffding, ``Probability inequalities for sums of bounded random
  variables,'' \emph{Journal of the American Statistical Association}, vol.~58,
  no. 301, pp. 13--30, 1963.

\bibitem{Serfling1974}
R.~J. Serfling, ``Probability inequalities for the sum in sampling without
  replacement,'' \emph{The Annals of Statistics}, vol.~2, no.~1, pp. 39--48,
  1974.

\bibitem{DupuisFawziRenner16EAT}
\BIBentryALTinterwordspacing
F.~Dupuis, O.~Fawzi, and R.~Renner, ``Entropy accumulation,'' arXiv:1607.01796,
  2016. [Online]. Available: \url{https://arxiv.org/abs/1607.01796}
\BIBentrySTDinterwordspacing

\end{thebibliography}

\end{document}